\documentclass[prd,nofootinbib,twocolumn,english]{revtex4}
\usepackage{amsmath}
\usepackage{amsfonts,color}
\usepackage{tensor}
\usepackage{amssymb,float}
\usepackage[utf8]{inputenc}
\usepackage{color}
\usepackage[unicode=true,pdfusetitle,
 bookmarks=true,bookmarksnumbered=true,bookmarksopen=true,bookmarksopenlevel=3,
 breaklinks=false,pdfborder={0 0 1},backref=false,colorlinks=true]
 {hyperref}
\usepackage{enumitem}
\usepackage{graphicx}
\usepackage{appendix}
\usepackage{hyperref}
\usepackage{lipsum}

\makeatletter
\long\def\dddddot#1{%
  {\mathop {#1}\limits ^{\vbox to-1.4\ex@ {\kern -\tw@ \ex@ \hbox {\normalfont .....}\vss }}}%
}
\long\def\multidots#1#2{%
  \count@=0
  {{\mathop {#2}\limits ^{\vbox to-1.4\ex@ {\kern -\tw@ \ex@ \hbox {\normalfont %
  \loop%
  \ifnum#1>\count@%
  .%
  \advance\count@ by1%
  \repeat%
  }\vss }}}}%
}
\makeatother

\begin{document}

\title{Spherically symmetric solutions with dust and radiation sources in Schouten-Codazzi Gravity}

\author{Sebastián N\'ajera}
\email{najera.sebastian@ciencias.unam.mx}
\affiliation{Instituto de Ciencias F\'isicas, Universidad Nacional Aut\'onoma de M\'exico, 62210, Cuernavaca, Morelos.}

\author{Roberto A. Sussman}
\email{sussman@nucleares.unam.mx}
\affiliation{Instituto de Ciencias Nucleares, Universidad Nacional Aut\'onoma de M\'exico, 
Circuito Exterior C.U., A.P. 70-543, M\'exico D.F. 04510, M\'exico.}

\author{Juan Carlos Hidalgo}
\email{najera.sebastian@ciencias.unam.mx}
\affiliation{Instituto de Ciencias F\'isicas, Universidad Nacional Aut\'onoma de M\'exico, 62210, Cuernavaca, Morelos.}

\begin{abstract}
We examine in this paper exact solutions of Schouten-Codazzi gravity (SCG), a second order metric theory in which a Codazzi tensor is added to the Einstein field equations. We consider two spherically symmetric geometries that describe null radiation and inhomogeneous dust, corresponding to generalized Vaidya and Lema\^itre-Tolman-Bondi (LTB) spacetimes. The Codazzi tensor has to comply the Codazzi equations, this imposes restrictions on the geometry of spacetime that are determined up to two integration functions that naturally generate an effective ``cosmological constant'' and an additional degree of freedom. We show that for the Vaidya geometry, these Codazzi contributions modify the Misner-Sharp mass, while the LTB models have as an evolution equation a cubic polynomial whose coefficients depend on the Codazzi integration functions. This leads to a broader class of dynamics than those of General Relativity, and allows for an accelerated expansion and ever-expanding models without the introduction of ``dark energy''. Overall, our analysis suggests that Schouten-Codazzi gravity provides an appealing framework for applications in cosmology. 
\end{abstract}

\maketitle

\section{Introduction}

One of the principal motivations for studying modified theories of gravity is that standard cosmology cannot explain astrophysical and cosmological observations without introducing dark matter and dark energy components, whose physical origin remains unknown. This issue has motivated the study of modified theories of gravity in which at least part of the dark sector phenomenology could arise from geometric effects rather than from additional dark sources \cite{clifton2012modified, peebles2003cosmological}. 

Among available alternative gravity theories, we have proposed a second order theory denoted by ``Schouten-Codazzi gravity'' (SCG) that introduces a new geometric framework based on the Schouten tensor and the Codazzi condition \cite{sussman2024schouten}. 

SCG is based on a general symmetric Codazzi tensor endowed with the algebraic structure of second order curvature tensors (including the Ricci and Schouten tensors), thus  enhancing  the degrees of freedom of gravitational dynamics with input from a diverse set of gravitational sources and geometries. SCG field equations are constructed by combining the Schouten tensor with this generic (or ``prospective'') Codazzi tensor, keeping the same energy momentum tensor of GR as the only physical field source. 

The  degrees of freedom introduced by SCG at the Hubble scale of an FLRW spacetime have the potential to reproduce, not only  the cosmological constant (appearing naturally as a constant of integration), but a generic accelerated cosmic expansion that can fit predictions of  the $\Lambda$CDM model and/or dynamical dark energy models \cite{sussman2024schouten}. SCG also introduces non-trivial geometric degrees of freedom that modify the gravitational dynamics at different scales. We have already derived exact solutions in static spherical symmetry: a Schwarzschild-like vacuum solution and a perfect fluid solution with a modified  Tolman-Oppenheimer-Volkov  hydrostatic equilibrium condition. It is also important to examine the degrees of freedom of SCG on inhomogeneous and anisotropic spacetimes used in astrophysical contexts. 

In the  paper, we investigate the cosmological and astrophysical implications of SCG by finding exact solutions in different geometrical physical scenarios. In particular, we analyze the SCG field equations and show that these lead naturally to integration functions which mimic an effective ''cosmological constant'' contribution. These functions provide a geometric interpretation of accelerated expansion without the need of an extra dark energy component. We find exact solutions for Lema\^itre-Tolman-Bondi (LTB) and Vaidya metrics, describing inhomogeneous configurations, and for the latter case with energy fluxes. These models have proven to be useful in the study of modified gravity theories \cite{kobayashi2005vaidya,cai2008generalized, sharif2014lemaitre,sussman2017lemaitre, najera2018geometric, yu2020lemaitre,  schiavone2024signature}. Our results show that SCG extends the physical description of General Relativity and provides a gravitational theory that can mimic ``dark energy'' sources as geometric contributions. 

The section by section content of the paper is described as follows: Sec.~\ref{sec:SCG} provides a review of Schouten-Codazzi gravity, while Sec.~\ref{sec:Solving} presents a detailed procedure for solving the Schouten-Codazzi field equations by exploiting the Segr\'e classification of the underlying spacetime and, whenever present, its symmetries. In Secs.~\ref{sec:vaidya}-\ref{sec:LTB}, we derive the generalized Vaidya and LTB spacetimes in detail. Finally, Sec.~\ref{sec:remarks} provides discussion and closing remarks.

\section{Schouten-Codazzi Gravity}\label{sec:SCG}

Schouten-Codazzi Gravity (SCG) is a second order metric theory of gravity. It is constructed from the Schouten tensor, the energy momentum tensor, and an additional symmetric tensor that satisfies the Codazzi differential condition. SCG was originally proposed \cite{sussman2024schouten} as an alternative geometric theory of gravity inspired by the structure of the Cotton Gravity framework developed by Mantica and Molinari \cite{mantica2023codazzi, mantica2024covariant}. SCG is completely independent from Cotton Gravity and avoids the issues associated with the Cotton tensor and the higher order field equations of Cotton Gravity.  

SCG is constructed with the following fundamental tensors: the Schouten tensor 
\begin{equation*}
S_{ab}= R_{ab}-\frac16 Rg_{ab},
\end{equation*}
and the trace modified energy momentum tensor
\begin{equation*}
\hat{T}_{ab}=T_{ab}-\frac13 Tg_{ab},
\end{equation*}
where $R_{ab}$ and $T_{ab}$ denote the Ricci and energy momentum tensors, respectively. The fundamental tensor that makes SCG differ from Cotton Gravity is a symmetric tensor $F_{ab}$, henceforth named a \emph{Codazzi tensor}, that must satisfy the Codazzi differential condition
\begin{equation}
\nabla_a F_{bc}-\nabla_c F_{ba}=0.\label{eq:codeq}
\end{equation}
The algebraic form of this Codazzi tensor will be constrained by the invariant properties of the spacetime. This tensor represents a deviation from GR and provides the extra degrees of freedom that 
This tensor contains the deviation from GR and the additional geometric degrees of freedom that characterize the theory. As stated above, $F_{ab}$ is not arbitrary. One must construct the tensor based on two basic requirements. First, it has to satisfy the Codazzi differential condition Eq.~(\ref{eq:codeq}). Secondly, the algebraic structure of this tensor must be compatible with the invariant properties of the spacetime \cite{sussman2024schouten, santos1995segre, stephani2009exact}. This requirement follows from the fact that all second order symmetric tensors in a four dimensional Lorentzian manifold have a common algebraic classification in terms of their eigenvectors and eigenvalues. Therefore, the tensor $F_{ab}$ is determined by the same invariant structure that defines the Ricci tensor and/or the energy momentum tensor, cf. Sec.~\ref{sec:Solving}. The choice of the Schouten and Codazzi tensors as the fundamental geometric objets of SCG is motivated by the success of Cotton Gravity in providing a theory that can explain the accelerated expansion of the Universe as well as the galactic rotation curves \cite{harada2022cotton}, and also by the inherent problems that the theory possesses such as having every conformally flat spacetime as a trivial solution. Rather than using a third order Cotton tensor obtained from derivatives of the Schouten tensor, SCG incorporates a Codazzi tensor directly into the field equations as an independent geometric object. Hence, the theory preserves the geometric motivation provided by Mantica and Molinari while remaining second order in the metric and avoiding many of the difficulties commonly encountered in higher order gravitational theories. 

The theory also has a well defined correspondence limit to GR. If the Codazzi tensor vanishes,
\begin{equation*}
F_{ab}=0,
\end{equation*}
the field equations reduce to 
\begin{equation*}
S_{ab} = 8\pi \hat{T}_{ab},
\end{equation*}
which are equivalent to Einstein's equations. 

\section{Solving Schouten-Codazzi equations}\label{sec:Solving}

The Schouten-Codazzi field equations (SCFE) can be written in a Ricci and Einstein "effective" form \cite{sussman2024schouten}
\begin{eqnarray}
R_{ab}+F_{ab}+\frac12 F g_{ab} &=& 8\pi \left [ T_{ab}-\frac12 T g_{ab}\right],\label{eq:SCR}\\
G_{ab}+F_{ab}-F g_{ab} &=& 8\pi T_{ab},\label{eq:SCG}
\end{eqnarray}
where $F=g^{cd}F_{cd}$ and $T=g^{cd}T_{cd}$ denote the traces of the Codazzi and energy momentum tensors, respectively.

The field equations can be formally rewritten in an ``Einstein like'' form by transferring the terms involving $F_{ab}$ to the matter sector. However, in SCG the tensor $F_{ab}$ is regarded as part of the geometric sector rather than as an effective matter content. This distinction is important because $F_{ab}$ is not an arbitrary matter source, it must satisfy the Codazzi differential condition Eq.~(\ref{eq:codeq}) which strongly restricts its functional form. If this tensor were to be interpreted as an effective energy momentum tensor, only a very special subset of possible sources would be allowed by this condition. As we shall see later, such admissible configurations are not generic matter distributions, but naturally lead to effective contributions with a dark energy like behavior. This is why it is more natural to think of $F_{ab}$ as part of the geometric sector rather than as an additional matter contribution. 

Furthermore, the algebraic form of this tensor must relate to the invariant symmetries and scalars that characterize the considered spacetime. Since $F_{ab}$ is a symmetric rank two tensor, it belongs to the same algebraic class as the Ricci and energy momentum tensors, and can therefore be characterized through the Segre classification. This classification is obtained from the eigenvalue problem
\begin{equation*}
	\tensor{A}{^a _b} v^b=\lambda v^a,
\end{equation*}
where $A_{ab}$ is an arbitrary symmetric tensor \cite{hall1976classification, stephani2009exact, santos1995segre}. The Segre type gives the multiplicity of the eigenvalues together with the causal character of the corresponding eigenvector. Hence it gives a coordinate independent characterization of the invariant structure of the tensor. 

Therefore, the basic assumption used in SCG is that the Codazzi tensor, $F_{ab}$, must be concomitant with the already existing geometric structure in the spacetime \cite{sussman2024schouten}. In practice, this means that $F_{ab}$ is chosen to have the same eigenvector structure as the Ricci tensor, or the energy momentum tensor. Once the Segre type of these tensors has been identified, it is possible to find an invariant tetrad in terms of which a canonical form for $F_{ab}$ can be written. The eigenvectors are inherited from the spacetime geometry, while the eigenvalues remain undetermined functions which are fixed by the Codazzi condition. 

The Segre classification does not uniquely determine the tensor $F_{ab}$, it only fixes the invariant algebraic structure by using the eigenvector decomposition and the degeneracies in the eigenvalues of each Segre classification, which are well known for every class of spacetime \cite{hall1976classification, stephani2009exact}. The eigenvalues will act as independent functions, which will be fully determined through the Codazzi restriction.  Therefore, the algebraic sector of the theory is given by the Segre classification, and the differential structure of SCG is given by the Codazzi equations with the SCG field equations. The explicit form of $F_{ab}$ depends on the Segre type of the considered spacetime. For example, if the Ricci tensor or the energy momentum tensor are of Segre type $[(11),1,1]$, then a compatible Codazzi tensor has the form
\begin{equation*}
F_{ab} = F_0 u_a u_b + F_1(x_a x_b + y_a y_b) + F_2 z_a z_b,
\end{equation*}
where $\{ u^a, x^a, y^a, z^a\}$ is an orthonormal tetrad adapted to the eigenvectors of the Ricci or the energy momentum tensors. Notice that the repeated eigenvalue $F_1$ corresponds to the degenerate two-dimensional eigenspace, in accordance with the Segre type. As stated previously, the decomposition of any symmetric 2-rank tensor of any Segre type given an invariant tetrad is well known. In general, these eigenvalues may depend on all spacetime coordinates. However, in spacetimes that admit symmetries their functional dependence has to be compatible with the corresponding isometries. For example, in a spherically symmetric spacetime with an orthonormal tetrad in spherical coordinates the eigenvalues will be functions of $(t,r)$. 

Once the algebraic form of $F_{ab}$ is specified, the Codazzi condition (\ref{eq:codeq}) gives a system of differential equations for the eigenvalue functions $F_i$. The resulting system will be coupled to the SCFE, both systems have to be solved simultaneosly. Hence, the Codazzi condition plays a double role: it restricts the possible forms of $F_{ab}$ and gives the differential relations to determine its components. 

The procedure for solving the SCFE can be summarized as follows. We first determine the Segre type of the Ricci or energy momentum tensor of the spacetime of under consideration. Secondly, we construct a tensor $F_{ab}$ with the same eigenvector structure as the tensor used to analyze the Segre type. Third, the Codazzi conditions are imposed to obtain a set of differential equations for the eigenvalue functions $F_i$. Finally, these equations are solved simultaneously with the SCFE. 

This formalism offers a systematic method for constructing admissible Codazzi tensors from the invariant algebraic properties of a spacetime. In the following sections, this method will be applied to several geometries of physical interest. The Segre type of the energy-momentum tensor will be determined in each case, a compatible Codazzi tensor will be constructed, and the corresponding Codazzi equations will be solved along with the SCFE. 

\section{Vaidya solutions in Schouten-Codazzi Gravity}\label{sec:vaidya}

In GR, the Vaidya spacetime serves as a natural generalization of the Schwarzschild geometry so that it can be applied to physical processes involving radiation \cite{stephani2009exact, vaidya1951nonstatic}. The Vaidya geometry describes the spherically symmetric gravitational field produced by a null fluid, and can be used to study systems in which energy is transferred radially through radiation. The Vaidya geometry has been used as a model for radiating compact objects, gravitational collapse, and black hole processes \cite{santos1985non, maharaj2012radiating, hiscock1981models}. Since there is a non-zero energy flux, the Vaidya spacetime becomes particularly relevant for testing modified theories of gravity \cite{kobayashi2005vaidya, cai2008generalized, rudra2025gravitational, gurses2025generalized}. The null radiation source has a distinctive algebraic structure that differs from the structure produced by a perfect fluid, providing constraints on the possible form of additional geometric tensors. 

The Vaidya spacetime is characterized by a matter content in the form of a purely radiative null fluid,
\begin{equation}
	T_{ab} = \nu l_a l_b,\label{Tabvai}  
\end{equation}
where $\nu=\nu(u,r)$ is the radiation density and $l^a$ is a null vector corresponding to the direction of the radiation flux. This type of source posseses a non trivial algebraic structure making it a suitable configuration for studying the role of the Codazzi tensor in SCG. In particular, the spherical symmetry of the spacetime and the algebraic structure related to the null fluid constrains the algebraic form of $F_{ab}$. 

We consider a spherical symmetric metric expressed in null coordinates
\begin{eqnarray*}
ds^2 &=& -f(u,r)du^2+2dudr +r^2(d\theta^2+\sin^2\theta d\phi^2),
\end{eqnarray*} 
 with the energy momentum tensor given by Eq.~(\ref{Tabvai}). The null vector associated with the radiation field is chosen as 
\begin{equation*}
l^a=g^{a b}\partial_b u = \tensor{\delta}{^a_r}, \qquad l_a dx^a = du.
\end{equation*} 
We complete $\{l^a\}$ to a half null tetrad $\{\mathbf x, \mathbf y, \mathbf k,\mathbf l\}$ that satisfies $k_al^a=-1$, $k_a k^a=l_a l^a=0$, $x_ax^a = y_a y^a=1$ by taking 
\begin{equation*}
	k_a dx^a =\frac{f}{2}du-dr, \quad x_a dx^a=rd\theta, \quad y_adx^a=r\sin\theta d\phi,
\end{equation*} 
whose corresponding vectors are 
\begin{eqnarray*}
l^a\partial_a&=&\partial_r,\qquad	k^a \partial_a =-\partial_u-\frac{f}{2}\partial_r, \\
x^a \partial_a&=&\frac{1}{r}\partial_\theta, \qquad y^a\partial_a=\frac{1}{r\sin\theta} \partial_\phi.
\end{eqnarray*} 
This type of spacetime is Petrov type D and the Segr\'e Type of the energy momentum tensor is \{(11),2\}, which as an algebraic decomposition of the energy momentum tensor in the half null tetrad as 
\begin{equation*}
T_{ab}=\lambda_1 x_a x_b + \lambda y_a y_b -2 \lambda_3 k_{(a} l_{b)}+l_a l_b,	
\end{equation*}
where $\lambda_i$ are the eigenvalues. The energy momentum tensor and the Codazzi tensor must have the same algebraic structure. Hence, we consider the algebraic decomposition 
\begin{equation}\label{eq:decomF}
F_{ab}=-2 F_1 k_{(a} l_{b)}+F_2 l_a l_b+F_3 x_a x_b + F_4 y_a y_b ,
\end{equation}
where the functions $F_i$ depend on $(u,r)$. The Codazzi differential condition Eq.~(\ref{eq:codeq}) implies the following system of equations
\begin{eqnarray}
F_3&=&F_4,\\
F_{1,r}&=&0,\label{eq:deflamu}\\
F_2&=&rF_{3,u},\label{eq:angcod}\\
F_1&=&(rF_3)_{,r}.\label{eq:radcod}
\end{eqnarray}
The remaining nonvanishing component of the Codazzi equations,
$$F_{1,u}+fF_{1,r}=F_{2,r},$$
is satisfied by this system since
$$F_{1,u}=\left[(rF_3)_{,r}\right]_{,u}=(rF_{3,u})_{,r}=F_{2,r}.$$
These equations determine the eigenvalues $F_i$ of the Codazzi tensor. Eq.~(\ref{eq:deflamu}) can be integrated directly to 
\begin{equation*}
F_1=\lambda (u),	
\end{equation*} 
where $\lambda(u)$ is an arbitrary function of the null coordinate.  
Substituting in Eq.~(\ref{eq:radcod}) and integrating yields
\begin{equation*}
	F_3=\lambda+\frac{\gamma}{r},
\end{equation*}
where $\gamma (u)$ is a second integration function. Now Eq.~(\ref{eq:angcod})
 can be solved as 
\begin{equation*}
F_2=r\lambda_{,u} +\gamma_{,u}.
 \end{equation*}
 
Substituting this result into the SCFE gives 
\begin{eqnarray}
r\lambda_{,u}+\gamma_{,u}-\frac{f_{,u}}{r}+&&\nonumber\\
f\left(3\lambda+\frac{2\gamma}{r}-\frac{\frac{\partial}{\partial r}(r(f-1))}{r^2}\right)&=&\kappa \nu, \label{eq:SCEVuu}\\
\frac{\frac{\partial}{\partial r}(r(f-1))}{r^2}&=&3\lambda+\frac{2\gamma}{r}, \label{eq:SCEVur}\\
\frac{\partial}{\partial r}(f_{,r}r^2)&=&2r(3r\lambda+\gamma). \label{eq:SCEVthth}
\end{eqnarray}
Eq.~(\ref{eq:SCEVur}) can be directly integrated leading to the metric function
\begin{equation}
f = 1-\frac{2M}{r} + \gamma r +\lambda r^2,	\label{eq:vaiymetfun}
\end{equation} 
where $M(u)$ is an arbitrary integration function, while Eq.~(\ref{eq:SCEVuu}) takes the usual Vaidya form 
\begin{equation}
	\kappa \nu(u,r) = \frac{2M_{,u}}{r^2}, \label{eq:eqnu}
\end{equation} 
Notice that in the case of the advanced ingoing null coordinate convention, which corresponds to the cross term $+2dudr$, the null energy condition $\nu \geq 0$ implies that $M_{,u}\geq 0$. Moreover, the angular component of SCFE Eq.~(\ref{eq:SCEVthth}) is automatically satisfied. 

The energy momentum conservation $\tensor{T}{^a^b_; _b}=0$ implies
\begin{equation*}
\frac{\partial}{\partial r}(r^2\nu)=0.
\end{equation*}
This is directly satisfied by Eq.~(\ref{eq:eqnu}). Hence, the energy momentum conservation does not place additional restrictions on either $\gamma(u)$ or $\lambda(u)$. Even though $\gamma$ and $\lambda$ modify the geometry through the metric functions, they do not explicitly contribute to the null radiation energy density, is still determined entirely by the ''mass function'' $M(u)$ obtained as an integration function. 

The geometrical meaning of this ''mass function'' $M(u)$ is related to the Misner-Sharp mass, $M_{MS}$ \cite{misner1964relativistic, hayward1996gravitational}. In a spherically symmetric spacetime, the quasilocal mass $M_{MS}$ is defined by 
\begin{equation*}
1-\frac{2M_{MS}}{r}=g^{ab}\nabla_a r\nabla_b r.
\end{equation*}
Since, for the metric under consideration, $g^{rr}=f$ and $r$ is the areal radius 
\begin{equation*}
	M_{MS}(u,r)=\frac{r}{2}(1-f).
\end{equation*}
Substituting Eq.~(\ref{eq:vaiymetfun}) we obtain 
\begin{equation*}
M_{MS}(u,r) = M(u)-\frac12\gamma(u) r^2 -\frac12 \lambda(u)r^3.
\end{equation*}
Notice that this mass contains the usual Vaidya mass contribution together with additional geometric terms obtained though the Codazzi tensor. Considering Eq.~(\ref{eq:vaiymetfun}), the term proportional to $\lambda r^2$ has the same functional dependence as the cosmological contribution appearing in the AdS-Vaidya solution of GR \cite{mallett1985radiating, wang1999generalized, babichev2013black}. However, within SCG this behaviour emerges from the Codazzi tensor instead of introducing a cosmological constant directly in the field equations. The linear term $\gamma r$ represents a new contribution, also present in Cotton Gravity \cite{gurses2025generalized}. 

The Ricci curvature associated with it will also be affected by the Codazzi tensor. Taking the trace of the SCFE,  
\begin{equation*}
R=-3F=-12\lambda (u)-\frac{6\gamma(u)}{r}.
\end{equation*}
Therefore, a nonvanishing $\gamma(u)$ produces a scalar curvature singularity at $r=0$.  

The SCG Vaidya solution extends the Vaidya radiating solution of GR. The energy flux retains the usual functional dependence as a derivative of the mass function, while the additional Codazzi functions modify the geometry in a way that mimics a ''cosmological constant'' without introducing additional matter sources.

\section{Lema\^itre-Tolman-Bondi solutions in Schouten-Codazzi Gravity}\label{sec:LTB}

In GR, the simplest exact solutions with an inhomogeneous source are the spherically symmetric Lema\^itre-Tolman-Bondi (LTB) model \cite{stephani2009exact, ellis2012relativistic, krasinski1997inhomogeneous, bolejko2010structures}. These solutions spatially inhomogeneous extensions to the Friedmann-Lema\^itre-Robertson-Walker (FLRW) cosmology. The LTB spacetimes have been widely used in the study of gravitational collapse, cosmic voids, and the role of inhomogeneities in cosmological evolution \cite{garcia2008confronting, bolejko2011inhomogeneous, sussman2015spherical, krasinski2001structure}. Since LTB models have inhomogeneous matter distributions while preserving spherical symmetry, they provide a natural framework for studying whether modified gravitational theories can reproduce cosmological effects without additional ``dark fluids'' \cite{sharif2014lemaitre, sussman2017lemaitre, yu2020lemaitre, najera2022inhomogeneous}. 

To obtain the SCG generalization of LTB models, we consider a spherically symmetric metric in comoving coordinates with a dust source
\begin{eqnarray}
ds^2 &=& -dt^2 + A^2(t,r)dr^2+Y(t,r)^2(d\theta^2+\sin^2\theta d\phi^2),\nonumber\\
\label{eq:LTBmet}\\
T_{ab}&=& \rho u_a u_b,\label{Tab}    
       \end{eqnarray} 
where $\rho=\rho (t,r)$ and $u^a$ is the comoving $4$-velocity. This space-time is Petrov type D and the energy momentum is Segr\'e type \{(111),1\}, therefore this tensor has the algebraic decomposition
$$T_{ab}=\lambda_1 x_a x_b+\lambda_2 y_a y_b+\lambda_3 z_a z_b+\lambda_0 u_a u_b,$$
 where $\{\mathbf x, \mathbf y, \mathbf z,\mathbf u\}$ is an orthonormal tetrad. In our particular case, we consider the orthonormal tetrad associated with the spherical coordinates in Eq.~(\ref{eq:LTBmet}). This gives us the decomposition for the $F_{ab}$ tensor
\begin{equation*}\label{eq:FabLTB}
\tensor{F}{^a_b}=F_1 x^a x_b + F_2 y^a y_b + F_3 z^a z_b -F_0 u^a u_b, 
\end{equation*}
where the functions $F_i$ are functions of $(t,r)$ and $x^a=e_r \delta^a_r$, $y^a=e_\theta \delta^a_\theta$, $z^a=e_\phi \delta^a_\phi$. The Codazzi differential condition implies the following system of equations
\begin{eqnarray*}
F_2&=&F_3,\\
F_{0_r}&=&0\\
(Y(F_2-F_1))_{,r}&=&-YF_{1_{,r}},\\
(Y(F_0-F_2))_{,t}&=&YF_{0_{,t}},\\
(A(F_0-F_1))_{,t}&=&AF_{0_{,t}}.
\end{eqnarray*}
The first of these equations reproduces the degeneracy of the two angular eigendirections required by spherical symmetry. 

The mixed $tr$ component of the SCFE coincides with the corresponding Einstein field equation, 
$$AY_{,tr}=Y_{,r}A_{,t},$$ 
which can be integrated to give,
\begin{equation*}
	A=\frac{Y_{,r}}{\sqrt{1-K}},
\end{equation*}
where $K=K(r)$ is a function related to the spatial curvature of the LTB geometry \cite{krasinski1997inhomogeneous}. 

Substituting this result into the previous system of equations, the Codazzi conditions can be rewritten as 
\begin{eqnarray}
F_2&=&F_3,\\
F_{0_r}&=&0\label{eq:defC0}\\
(Y(F_2-F_1))_{,r}&=&-YF_{1_{,r}},\label{eq22}\\
(Y(F_0-F_2))_{,t}&=&YF_{0_{,t}},\label{eq62}\\
(Y_{,r}(F_0-F_1))_{,t}&=&Y_{,r}F_{0_{,t}}.\label{eq52}
\end{eqnarray}
The remaining SCFE can then be expressed as 
\begin{eqnarray}
\kappa \rho -F_1-2F_2&=&\frac{\frac{\partial}{\partial r}\left(Y\left(Y_{,t}^2 +K\right)\right)}{Y_{,r}Y^2},\label{eq:SCEtt}\\
F_0+2F_2&=&-\frac{\frac{\partial}{\partial t}\left(Y\left(Y_{,t}^2 +K\right)\right)}{Y_{,t}Y^2},\label{eq:SCErr}\\
F_0+F_1+F_2 &=& -\frac{\frac{\partial}{\partial r}\left(\frac{\frac{\partial}{\partial t}\left(Y\left(Y_{,t}^2 +K\right)\right)}{Y_{,t}}\right)}{2Y_{,r}Y}.\label{eq:SCEthth}
\end{eqnarray}

The Codazzi equations determine the allowed functional form of the eigenvalues of $F_{ab}$. In particular, Eq.~(\ref{eq:defC0}) is easily integrated to $F_0=\lambda(t)$, while Eqs.~(\ref{eq62}), (\ref{eq52}) allow $F_1$ and $F_3$ to be written as quadratures 
\begin{eqnarray*}
	F_1 &=& \frac{\int \lambda Y_{,tr}dt + \xi}{Y_{,r}},\\
	F_3 &=& \frac{\int \lambda Y_{,t}dt + \gamma}{Y},
\end{eqnarray*}
where $\gamma,\xi$ are arbitrary functions of $r$\footnote{In the case $\rho =0$, one recovers the Schwarzschild model and $\gamma, \xi$ will be constant.} arising from the integration with respect to $t$. From the remaining Codazzi condition, Eq.~(\ref{eq22}), it is easy to show that $\xi=\gamma_{,r}$. \\

Therefore, the degrees of freedom of the Codazzi tensor are entirely encoded in the free functions $\lambda$ and $\gamma$, just as in the Vaidya and FLRW cases \cite{sussman2024schouten}. These functions characterize the additional geometric contribution of the Codazzi tensor and must be prescribed such that they obey initial conditions. Substituting the expressions obtained from the Codazzi equations into the SCFE, and integrating Eq.~(\ref{eq:SCErr}) we obtain the evolution equation for the areal radius $Y(t,r)$
\begin{equation}\label{eq:Yt2LTB}
	Y_{,t}^2 = -\frac{\int (2F_3 + \lambda)Y^2 Y_{,t} \, dt}{Y} +\frac{2M}{Y}-K =\frac{2M}{Y}-K -F_3Y^2, 
\end{equation}
where $M=M(r)$ is the generalized mass function and $K=K(r)$ is the curvature function. This equation is analogous to the Friedmann equation in FLRW models. The conservation equations $\tensor{T}{^a ^b_; _b}=0$ as well as Eqs.~(\ref{eq:SCErr})-(\ref{eq:SCEthth}) are then solved with Eq.~(\ref{eq:Yt2LTB}), while Eq.~(\ref{eq:SCEtt}) takes the form
\begin{equation}\label{eq:eqrhoLTBvar}
\kappa \rho =F_1+2F_3+\frac{2M_{,r}-(F_3Y^3)_{,r}}{Y_{,r}Y^2}.	
\end{equation}
For the purpose of establishing a connection with the cosmological constant, we shall consider the particular case in which $\lambda$ is constant. This choice yields 
\begin{eqnarray}
	F_1 &=& \lambda + \frac{\gamma_{,r}}{Y_{,r}},\nonumber\\
	F_3 &=& \lambda + \frac{\gamma}{Y},\nonumber\\
	Y_{,t}^2 &=& \frac{2M}{Y}-K- \gamma Y-\lambda Y^2,\label{eq:Yt2LTBcons}
\end{eqnarray}
while Eq.~(\ref{eq:eqrhoLTBvar}) reduces to the usual LTB dust density relation \cite{krasinski1997inhomogeneous}
\begin{equation*}
\kappa \rho = \frac{2M_{,r}}{Y_{,r}Y^2}.
\end{equation*}
Notice that identifying $\lambda=-\Lambda/3$ in Eq.~(\ref{eq:Yt2LTBcons}) and taking $\gamma=0$ recovers the standard $\Lambda$LTB cosmology. Therefore, the cosmological constant can be interpreted in SCG as an effective geometric term arising from the Codazzi sector.

As shown in \cite{najera2018geometric} the kinematic evolution of Eq.~\ref{eq:Yt2LTBcons} is governed by the zeroes of a cubic polynomial 
\begin{equation}
	Y_{,t}^2 = \frac{Q(Y)}{Y}, \qquad Q(Y) = 2M -KY - \gamma Y^2 +\frac{\Lambda}{3}Y^3,\label{eqQY}
\end{equation}
for different values of $K$, where we identified $\lambda=-\Lambda/3$. By differentiating Eq.~(\ref{eqQY}), we obtain
\begin{equation*}
	Y_{,tt} = -\frac{M}{Y^2}-\frac{\gamma}{2}+\frac{\Lambda}{3}Y^2,
\end{equation*}
therefore for $\gamma >0$ this function produces an additional deceleration, while $\gamma<0$ produces a positive acceleration contribution. 
 \begin{figure}[h]
\centering
\includegraphics[width=0.9\linewidth]{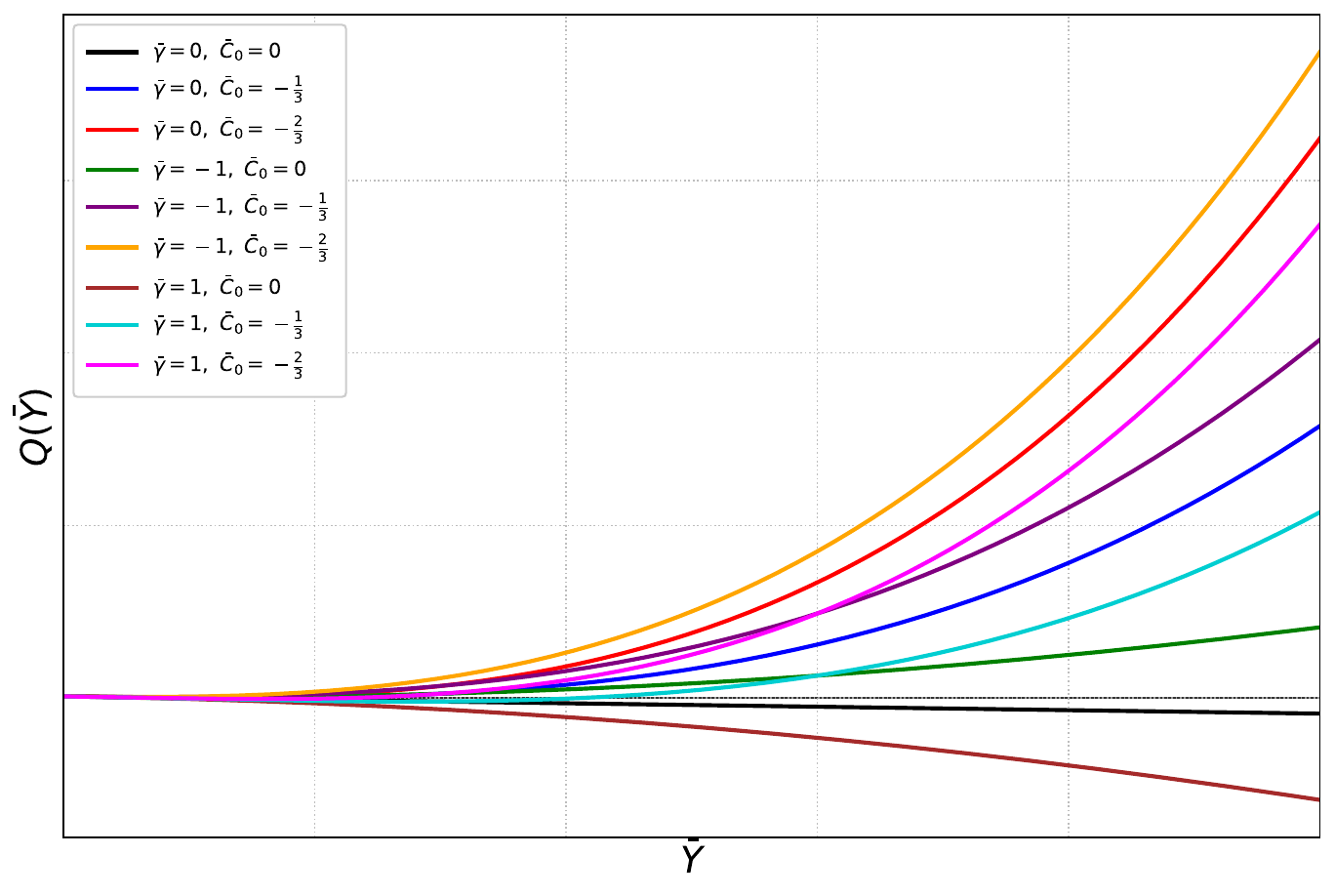}
\caption{$Q(\bar Y)$ in Eq.~(\ref{eqQY}), as a function of the dimensionless areal radius $\bar Y$ for representative LTB models with fixed values of the parameters $\bar M, K$, and different choices of $\bar\lambda, \bar\gamma$.}
\label{fig:fig1}
\end{figure}

To present numerical solutions in dimensionless form, we use the Hubble scale $H_0^{-1}$ and define $\bar t =H_0 t$, $\bar{Y} =H0 Y$, $\bar M=H_0M, \bar \gamma = \gamma /H_0$ and $\bar\Lambda = \Lambda/H_0^2$. In terms of these variables, the evolution equation Eq.~(\ref{eqQY}) retains the same algebraic form. In Fig.~\ref{fig:fig1} we illustrate the behaviour of the cubic polynomial $Q(\bar Y)$ for fixed positive values of $\bar M$ and an arbitrary positive curvature parameter $K>0$. A negative value of $\bar\gamma$ enhances the expansion, whereas a positive $\bar \gamma$ suppresses it, leading to a decelerated evolution. However, the dynamical effect of $\bar\gamma$ depends on $\bar Y$, while the $\bar\lambda$ term scales as $\bar Y^2$. Notice both of this contributions play the role of an effective ``cosmological constant''. The qualitative behaviour of these solutions is analogous to that reported in Ref.~\cite{najera2018geometric}.

In the particular case $\Lambda=0$, the turning points of a given comoving shell are determined by $Q(Y)=0$, which for $\gamma\neq 0$ possesses real zeros whenever 
\begin{equation*}
K^2 +8\gamma M\geq 0,	
\end{equation*}
where the equality corresponds to a double root. The corresponding roots are
$$ Y_\pm=\frac{-K\pm \sqrt{K^2+8M\gamma}}{2\gamma}.$$
However, only positive roots correspond to physical turning radii, while the allowed region must additionally satisfy $Q(Y)\geq 0$ for $Y>0$. 

Notice that whenever $\gamma <0$ this allows for a selection of free functions that allows ever expanding dust models, independently of the sign of K, but they do not occur for every choice of free functions.  

It is easy to check that whenever $\Lambda \neq 0$ an analogous analysis can be done. The detailed dynamical behaviour depends on the cubic discriminant and on the choice of free functions. The special case $\Lambda\neq 0, \gamma =0$ has been studied in detail in Ref.~\cite{najera2018geometric}.
 
  The above analysis generalizes the results of Ref.~\cite{najera2018geometric} by showing that the additional Codazzi degree of freedom, encoded in the parameter $\gamma$ enlarges the family of ever expanding LTB dust solutions. While GR already admits ever expanding configurations, Schouten Codazzi gravity provides a broader class of models in which the expansion history may be accelerated or decelerated depending on the sign of $\gamma$, in combination with the effective cosmological constant associated with $\lambda$.
  
  In the collapsing case, we can quantify the time for a collapsing shell to reach $Y=0$ as 
\begin{equation*}
  t_c(r)-t_0 = \int_0^{Y(0,r)}\frac{dY}{\sqrt{2M/Y-K-\gamma Y +\Lambda Y^2/3}},
  \end{equation*}
  where $t_c$ is the collapsing time and $t_0$ is the Big Bang time. \\
  
\begin{figure}[h]
\centering
\includegraphics[width=0.9\linewidth]{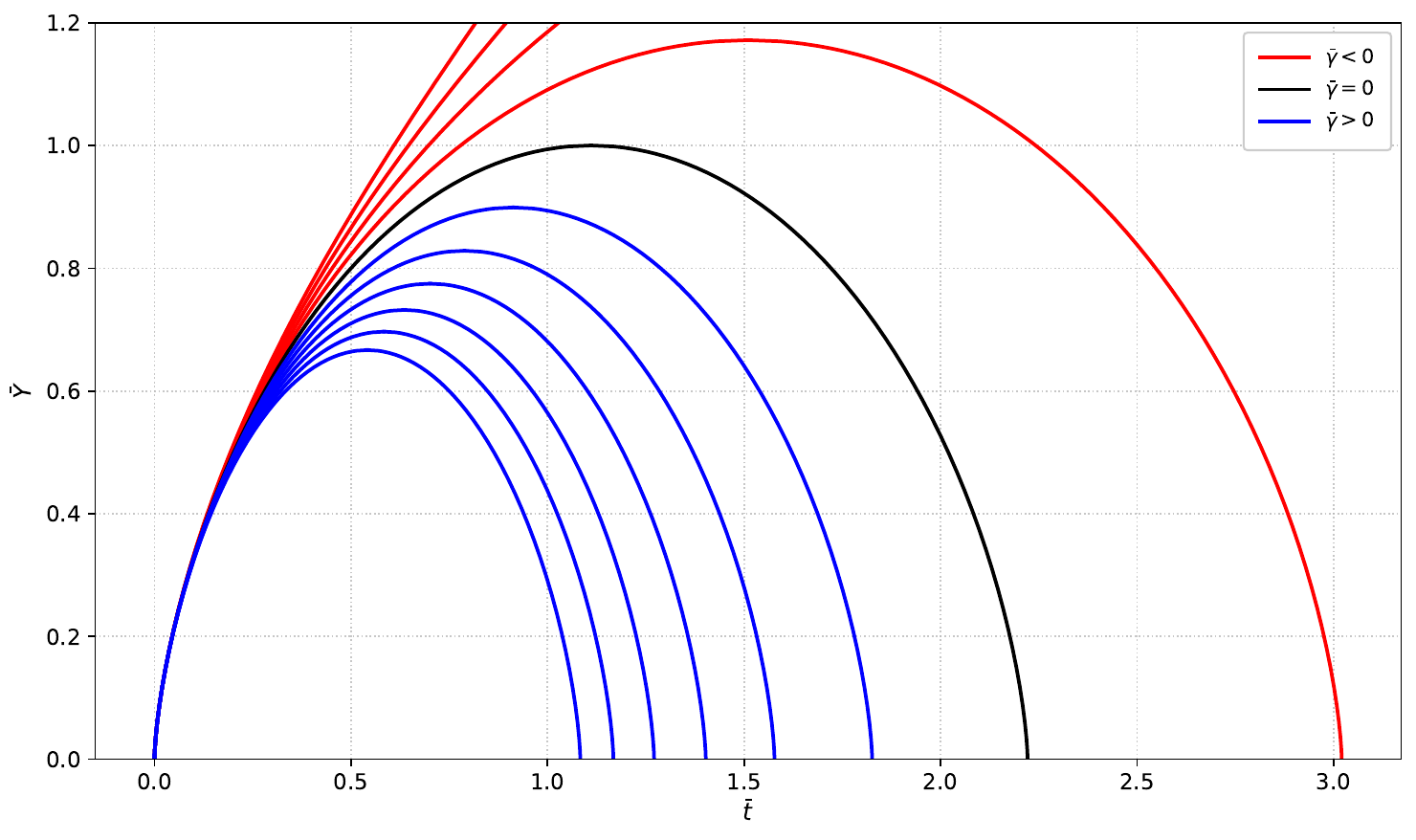}
\caption{Dimensionless areal radius $\bar Y$ as a function of the dimensionless time $t$ for representative LTB models with fixed values of the parameters $\bar M, K, \bar \lambda$ and different values of $\bar \gamma$. The shells in these models evolve according to Eq.~(\ref{eqQY}).}
\label{fig:fig2}
\end{figure}

  This expression, given a selection of free functions, provides a quantitative measure of the influence of the Codazzi sector's contribution to the collapse. For $\gamma >0$, the additional Codazzi contribution enhances the collapse, in contrast with the $\gamma <0$ case, which can favour ever-expanding solutions. The effect of $\bar\gamma$ on the collapse dynamics is illustrated in Fig.~\ref{fig:fig2} for fixed positive values of the parameters $\bar M$ and $K$. The black curve corresponds to $\bar\gamma=0$, which recovers the GR case. This shows that the time collapse can be modified using $\bar\gamma$ as a new parameter.

\section{Discussion and final remarks}\label{sec:remarks}

The purpose of this article is to analyze and present exact solutions in the framework of Schouten-Codazzi gravity and their potential as appealing models in cosmology. For the present paper, we focus on spherically symmetric geometries, for which we derive exact solutions describing null radiation and inhomogeneous dust, corresponding to generalized Vaidya and Lema\^itre-Tolman-Bondi (LTB) spacetimes. We have chosen these geometries as they have proven to be valuable spacetimes that generate physically relevant toy models in cosmology. 

The Codazzi equations determine the geometric sector up to integration functions that naturally generate a contribution analogous to a cosmological constant together with an additional degree of freedom. These functions allow for the existence of models which resemble dynamical dark energy, while remaining in a second-order theory, obtained directly through the field equations. 

 For the generalized Vaidya spacetimes, these geometric contributions modify the mass function while preserving the characteristic structure of the spacetime. The Codazzi sector extends the behaviour beyond that of a cosmological constant behaviour. This motivates further investigation of the geometry, the causal structure and the physical interpretation of the radiating solutions in Scouhten-Codazzi gravity. 
 
 In the LTB case, the integration functions arising from the Codazzi sector modify the evolution equation of the areal radius, which determines the turning points of the models. This equation is governed by a cubic polynomial whose coefficients depend on the Codazzi sector, leading to a broader class of cosmological dynamics than in GR. In particular, although ever expanding ``closed'' LTB dust models are already present in GR, Schouten-Codazzi gravity enlarges this family of solutions through an additional geometric degree of freedom, allowing a wider range of expansion histories characterized by an increased number of free parameters.
  
 Our results extend previous analyses obtained in the absence of this additional degree of freedom and illustrate how Schouten-Codazzi gravity admits exact solutions with rich phenomenology without introducing higher-order field equations. Nevertheless, these exact solutions require further investigation, such as the regularity of the spacetimes, the existence of shell-crossing singularities, and considering other physically admissible sources, such as imperfect fluids in the LTB case. Overall, our analysis suggests that Schouten-Codazzi gravity provides an appealing framework for applications in cosmology.

\begin{acknowledgments}
SN acknowledges support from a postdoctoral grant from Secretar\'ia de Ciencia, Humanidades, Tecnolog\'ia e Innovaci\'on SECIHTI. SN and JCH acknowledge support from PAPIIT-UNAM project No. IN114626 "Estudios de ANPs y contribuciones In-Kind a la Colaboraci\'on LSST-Mexico"
\end{acknowledgments}

\bibliographystyle{apsrev4-1}
\bibliography{references.bib}
\end{document}